\documentclass{llncs}
\usepackage{makeidx}  % allows for indexgeneration
\usepackage{color}
\usepackage{hyperref}
\usepackage{graphicx}
\usepackage{wrapfig}

\begin{document}
%
%\frontmatter          % for the preliminaries
%
\pagestyle{headings}  % switches on printing of running heads

\title{Power Consumption-based Detection of Sabotage Attacks in Additive Manufacturing}
\author{Samuel B. Moore\inst{1} \and
Jacob Gatlin\inst{1} \and
Sofia Belikovetsky\inst{2} \and
Mark Yampolskiy\inst{1} \and
Wayne E. King\inst{3} \and
Yuval Elovici\inst{2,4,5}
}
\institute{
University of South Alabama
\and
Ben-Gurion University
\and
Lawrence Livermore National Laboratory
\and
Cyber Security Research Center
\and
Singapore University of Technology and Design
}

\maketitle              % typeset the title of the contribution

\begin{abstract}

Additive Manufacturing (AM), a.k.a. 3D Printing, is increasingly used to manufacture functional parts of safety-critical systems.
AM's dependence on computerization raises the concern that the AM process can be tampered with, and a part's mechanical properties sabotaged. This can lead to the destruction of a system employing the sabotaged part, causing loss of life, financial damage, and reputation loss.
To address this threat, we propose a novel approach for detecting sabotage attacks.
Our approach is based on continuous monitoring of the current delivered to all actuators during the manufacturing process and detection of %anomalies compared to
deviations from
a provable benign process.
The proposed approach has numerous advantages: (i) it is non-invasive in a time-critical process, (ii) it can be retrofitted in legacy systems, and (iii) it is airgapped from the computerized components of the AM process, preventing simultaneous compromise.
Evaluation on a desktop 3D Printer detects all attacks involving a modification of X or Y motor movement, with false positives at 0\%.

\end{abstract}
%

% ==================================================================
% === INTRODUCTION
% ==================================================================
\section{Introduction}

Additive Manufacturing technology (AM), a.k.a. 3D Printing, is receiving immense attention due to the potential for improvements in product performance, decreased development times, and reduced costs. 
As compared to either traditional, ``subtractive'' manufacturing methods in which material is removed from a part via machining or the use of pre-fabricated dies (e.g. investment casting, die casting, injection molding) which necessitate substantial investments in capital equipment, longer lead times, and associated labor costs, AM allows for the production of components with minimum material waste, shorter design-to-production times, and economical, on-demand production of niche parts. 

These advantages enable a broad range of applications, ranging from models and prototypes up to functional parts of safety-critical systems. An example of the latter is the FAA-approved fuel nozzle for General Electric's LEAP jet engine~\cite{GE2015faa}. 
Further examples of components produced using AM techniques include medical implants~\cite{schwindling2015two,monroy2013study}, 
air ducts~\cite{khajavi2014additive}, and tooling~\cite{santos2006rapid}.
According to the \emph{Wohlers Report}~\cite{wohlers2016report}, a renowned annual survey of advances in AM, in 2015 the AM industry accounted for \$5.165 billion of revenue, with 32.5\% of all AM-manufactured objects used as functional parts. 
A study conducted by \emph{Ernst \& Young}~\cite{ey2016how} 
shows rapidly growing adoption of this technology worldwide. 
In the U.S. alone, 16\% of surveyed companies have experience with AM and another 16\% are considering adopting this technology in the future. The current world leader of AM adoption is Germany with 37\% of surveyed companies already using and a further 12\% considering AM. 

Because of AM's dependence on computerization, there is a growing concern that the AM process can be tampered with, in order to sabotage a part's mechanical properties. 
While several studies have sabotaged a part's mechanical properties in a lab~\cite{sturm2014cyber,yampolskiy2015security,zeltmann2016manufacturing}, a recent \emph{dr0wned} study~\cite{belikovetsky2017dr0wned} has proven experimentally that a complete sabotage attack is possible.

To address the emerging security threat, in this paper we propose and evaluate a novel solution for detecting sabotage attacks in AM. 
The proposed solution is based on a monitoring of current supplied to individual actuators and the detection of anomalies in this data. 
The proposed approach has numerous advantages: (i) it is non-invasive in a time-critical process (ii) it can be retrofitted in legacy systems, and (iii) it can be easily air-gapped from the computerized components involved in the AM process, 
increasing the difficulty of simultaneous compromise. 

The remainder of this paper is structured as follows.
After discussing related work in Section~\ref{sec:sota} and describing an attack model in Section~\ref{sec:threat}, we present details of the proposed solution in Section~\ref{sec:solution}.
We present an experimental evaluation of the proposed solution in Section~\ref{sec:evaluation}.
After discussing the applicability of the proposed solution to industrial-grade metal AM systems in Section~\ref{sec:discussion}, we conclude this paper with a short review and an outline of planned future work.

% ==================================================================
% === THREAT MODEL
% ==================================================================

\section{Related Work} 
\label{sec:sota}

While researchers have already discussed various threats in the context of AM, such as Intellectual Property (IP) and manufacturing legally prohibited objects, only sabotage attacks and proposed detection methods are of relevance to this paper.

Sturm et al., 2014~\cite{sturm2014cyber} demonstrated that a part's tensile strength can be degraded by introducing defects such as voids (internal cavities).
Zeltmann et al., 2016~\cite{zeltmann2016manufacturing} showed that similar results can be achieved by printing part of the structure with the contaminated material.
Belikovetsky et al, 2017~\cite{belikovetsky2017dr0wned} proposed degrading a part's fatigue life; the authors argue that the defect's size, geometry, and location are factors in the degradation. 
Yampolskiy et al, 2015~\cite{yampolskiy2015security} argued that the anisotropy intrinsic to 3D printed parts can be misused to degrade a part's quality, if an object is printed in the wrong orientation.
Zeltmann et al., 2016~\cite{zeltmann2016manufacturing} have experimentally shown the impact of this attack on a part's tensile strength, using 90 and 45 degree rotations of the printed model.
Chhetri et al., 2016~\cite{chhetri2016kcad} introduced a skew along one of the build axes as an attack.
Moore et al., 2016~\cite{moore2017implications} modified the amount of extruded source material to compromise the printed object's  geometry.
Pope et al., 2016~\cite{pope2016hazard} identified that indirect manipulations like the modification of network command timing and energy supply interruptions can be potential means of sabotaging a part. 
Yampolskiy et al., 2015~\cite{yampolskiy2015security} discussed various metal AM process parameters whose manipulation can sabotage a part's quality; for the powder bed fusion (PBF) process, the identified parameters include heat source energy, scanning strategy, layer thickness, source material properties like powder size and form, etc.
Slaughter et al., 2017\cite{slaughter2017ensure} has shown that an indirect sabotage attack is possible via a compromised \emph{in-situ} infrared thermography; authors evaluate identified attacks on a metal 3D printer that employs the PBF process. 
Yampolskiy et al., 2016~\cite{yampolskiy2016using} argued that in the case of metal AM, manipulations of manufacturing parameters can not only sabotage a part's quality, but also damage the AM machine, or lead to the contamination of its environment.

Several publications present methods for detecting sabotage attacks.
Chhetri et al., 2016~\cite{chhetri2016kcad} used the acoustic side-channel inherent to the FDM process to detect tampering with a 3D printed object; the authors report that the detection rate of object modifications is 77.45\%.
Strum et al., 2017~\cite{strum2017insitu} proposed an impedance-based monitoring method. The authors physically coupled a piezoceramic (PZT) sensor to the part being fabricated and measure the electrical impedance of the PZT. These impedance measurements can be directly linked to the mechanical impedance of the part, assisting in detecting in-situ defects of part mass and stiffness. 
Two further papers built upon the cross-domain attack notion introduced in Yampolskiy et al., 2013~\cite{yampolskiy2013taxonomy}, and propose a notion of cross-domain attack detection.
Chhetri et al., 2017~\cite{chhetri2017cross} demonstrated the flow of information between the cyber and physical domains and how this information can be used for performing cross-domain security analysis. By estimating this relationship, the model can be used for the detection of new cross-domain attack models and attack detection techniques. 
Wu et al., 2017~\cite{wu2017detecting} leverage machine learning methods to detect cyber-attacks in the manufacturing process. The authors have used vision and acoustics as the data sources for machine learning algorithms and were able to detect anomalies with high accuracy (96.1\% and 91.1\% respectively). 

% ==================================================================
% === THREAT MODEL
% ==================================================================

\section{Threat Model}
\label{sec:threat}

3D printed objects generally begin as digital 3D models, stored on a computer. The most common file format remains \emph{.STL}, with emerging file formats like \emph{.AMF} or \emph{.3MF} offering better accuracy and additional features like color.
For a 3D printing job, the 3D model is first ``sliced'' by a dedicated software (like \emph{Slic3r}) into layers; for each layer a toolpath is generated that defines exactly what actuators (motors and similar 3D printer components) should act and in what sequence. 
Afterwards, the PC communicates to the 3D printer either the whole toolpath or individual commands (for desktop 3D printers, usually in \emph{G-code}). 
For the communication either network protocols or a USB connection is utilized.
All commands are then interpreted and executed by the firmware installed on the 3D Printer. 
If a command requests motion from an analog actuator, the firmware ``translates'' it to the actuator's input power supply characteristics, such as frequency and voltage. 

For sabotage attacks, researchers have compromised various elements of the outlined 3D printing workflow.
Moore et al., 2016 identified numerous vulnerabilities in software, firmware, and communication protocols often employed in desktop 3D printing niche.
Belikovetsky et al., 2016~\cite{belikovetsky2017dr0wned} used a phishing attack to enable remote access to the controller PC.
Sturm et al., 2014~\cite{sturm2014cyber} used malware running on the controller PC to automatically modify STL files.
Do et al., 2016~\cite{do2016data} exploited weaknesses in the communication protocol between the controller PC and a 3D printer and were able to cancel a print or submit a new job.
Moore et al., 2017~\cite{moore2017implications} presented a wide range of attacks possible through 3D Printer firmware compromise.

\begin{figure}[tbp]%[htbp]
	\centering
		\includegraphics[width=0.95\textwidth]{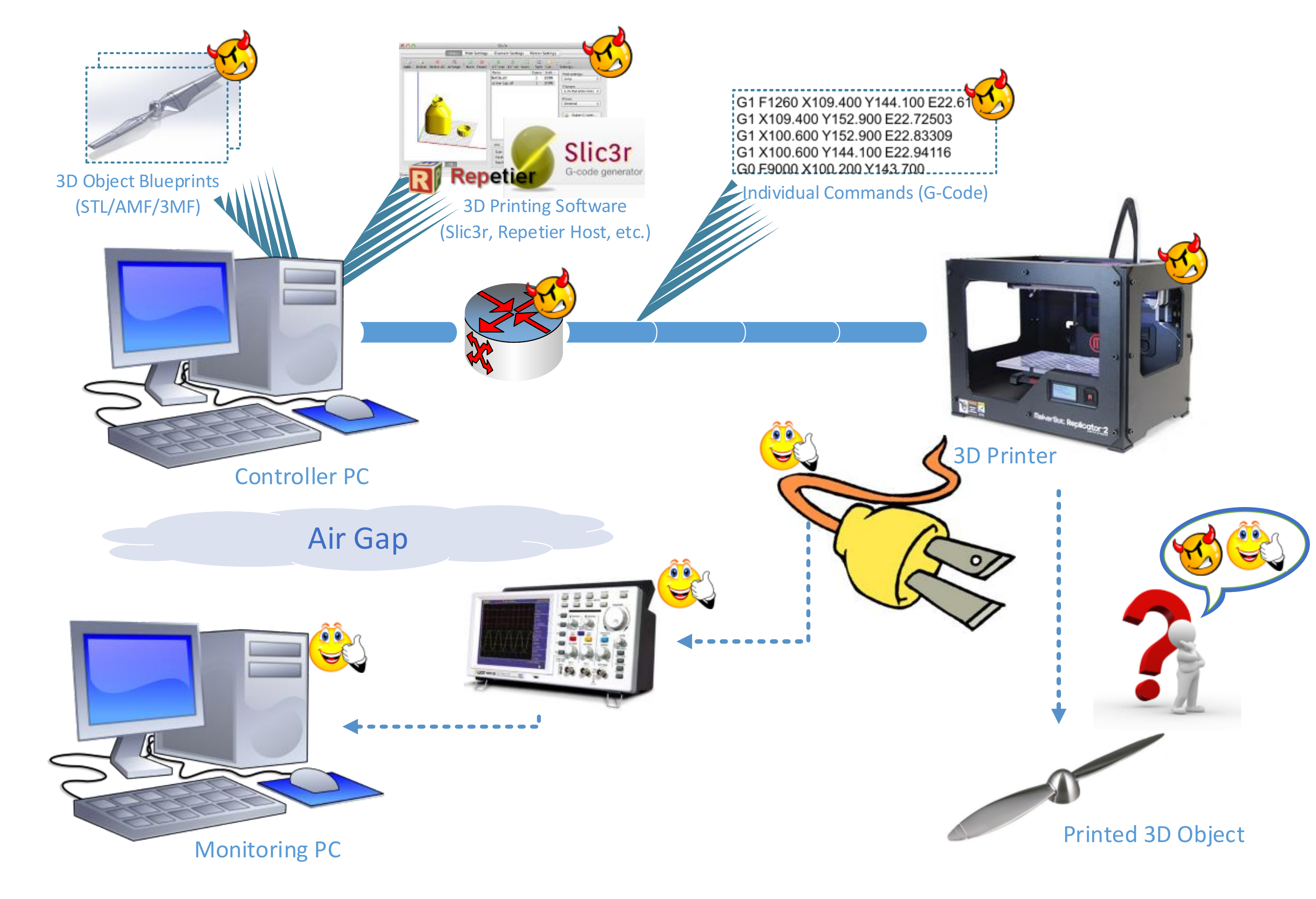}
	\caption{Considered Threat Model}
	\label{fig:ThreatModel}
\end{figure}

We consider the following threat model (summarized in Figure~\ref{fig:ThreatModel}):

\begin{itemize}
	\item As a lesson learned from the demonstrated attacks, we assume that any computerized element in the AM workflow (controller PC, 3D printer, computer network) can be compromised. 
	\item Analog actuators (e.g., stepper motors) cannot be compromised via cyber means. We assume that actuators behaving according to characteristics of analog input (such as Variable Frequency Drives, VFD) are not compromised. 
	\item Destructive testing can provide high-confidence data about mechanical characteristics and geometry of a manufactured 3D object. This information will be used to validate that the 3D printing process has not been tampered with.
	This approach is similar to the one used by Agrawal et al., 2007~\cite{agrawal2007trojan} for detecting hardware Trojans using IC fingerprinting.
	\item Our detection system (that includes induction probes, oscilloscope, and a monitoring PC for data analysis) is air-gapped from the manufacturing environment and is not compromised. 
	Results from destructive testing can be used to (manually) confirm the benign nature of a recorded manufacturing process.
	\item Electrical connections (of power monitoring system and of 3D Printing environment) are not physically tampered with, so that they are identical during both unaltered and maliciously modified manufacturing. 
\end{itemize}

% ==================================================================
% === THREAT MODEL
% ==================================================================

\section{Proposed Solution}
\label{sec:solution}

In this section, we outline the proposed approach, and provide details on how---according to our proposal---power supply signatures should be generated and compared.

\subsection{Considered Approach}

Our approach hinges on the direct, causative relation between the amplitude and frequency of current delivered to a motor and the motor's rotation. Any G-code move command specifies an (X,Y,Z) position to move the extruder head to, a speed to move at, and a path to get there. When these commands are received by the printer, the firmware translates them into a series of motor activations. The firmware communicates with the on-board motor controllers, which deliver current at a given frequency and amplitude to actuate the motors.

This translation from G-code, to motor activations, to current, is fully deterministic. Moreover, a set of delivered current over time for each motor will always result in the same printed object. The current in each motor is therefore the penultimate representation of the object, which began as a 3D model file. While various factors, including the mechanical arrangement of the motors, the filament, and the temperature of the extruder and bed, may influence the translation from current to physical object, this analog representation is not alterable by cyber means. 

The current passing through a wire may be measured, nonintrusively, using a current transformer. It can be sampled by an oscilloscope or digitizer; the sequence of samples for each motor circuit across the duration of a print is a measurement of the motor-current representation of the printed object.

Each trace in this representation is a series of (time, amplitude) data. If we align traces at some consistent starting point, e.g., the beginning of the first print layer, we should see the following: for prints with identical G-code, the traces will be equivalent, with periods of activity occuring at the same time and having the same frequency, with some level of desync due to the limited precision of the equipment. For prints with wholly different G-code, any alignment of the traces will be accidental and temporary; the difference in amplitude at any given time will vary from 0 to the full amplitude of the signal.

It follows from the above that prints with identical G-code, apart from some number of malicious modifications, will behave like this: up until the first modified command, the traces will display the same average deviation from each other as a comparison of two unmodified prints. When the first altered command is reached, the deviation will begin to vary across the full amplitude range. Depending on the duration of the command, the traces could either re-sync, or continue to produce the full range of deviations.

We propose an attack detection method based on the comparison of motor current traces. In brief, the method requires generating traces from several known-good prints, collecting the current traces of subsequent prints of the same object on that printer, and comparing the captured traces against the known-good traces. Modifications are identified by deviations between the captured and known-good traces being substantially different from the standard deviation. In the remainder of this section, we describe the individual stages in more detail.

\subsection{Trace Generation}

For a single combination of G-code and printer, a set of traces must be captured for each motor (or other actuator) involved in the printing process. The sample rate of the trace must be above the Nyquist rate, i.e., at least double the rate of the highest non-noise frequency component in the signal. The size of the trace set is determined by the statistical power of the method. In general, it will depend on the measurement error, standard deviation between ``normal'' traces, expected deviation of a malicious trace, and the acceptable false positive and negative rates.

An assumption of the method, as in the attack model, is that the printed objects can be physically verified. If it is impossible to verify every required property from a single object, it is sufficient to verify them from the same pool of objects created while capturing the known-good trace data.

When printing the potentially compromised object, a trace for each actuator must be captured, meeting the same standards as the known-good traces.

\subsection{Comparison of Traces}

Before comparing the traces, they must be time-aligned and preprocessed. Time-alignment can be done by pattern matching software, but is more easily achieved by a consistent hardware signal from the printer. Preprocessing should involve smoothing the traces to minimize the impact of sampling noise and error on the comparison. The smoothing method and strength must preserve the meaningful components of the signal, i.e., those that strongly impact the operation of the motor or actuator. In the case of multiphase AC motors, these would be the primary frequency components and their harmonics.

For each captured trace, compute the deviation across equivalent sample points on a known-good trace from the same motor.  If the captured trace does not correspond to unmodified G-code, it will produce a deviation over time similar to the standard deviation between known-good traces from that motor.

If it does correspond to modified G-code, the deviation will fluctuate across a much larger range starting at the first modified command. The detection threshold can be set in several ways; a simple version places a boundary above the highest peak in the known-good deviation, and detects a print as malicious if a certain number of samples have a larger deviation.

It may be useful to normalize the deviation of the captured trace to the standard deviation, or to produce a measure of accumulated deviation over time. These and other analysis methods are discussed in Future Work.

% ==================================================================
% === EXPERIMENTAL EVALUATION
% ==================================================================

\section {Experimental Evaluation}
\label{sec:evaluation}

In this section, we first present our experimental setup and describe the experiments performed.
We then present and analyze our results.

\subsection{Experimental Setup}

\begin{figure}[tbp]%[htbp]
	\centering
		\includegraphics[width=1.00\textwidth]{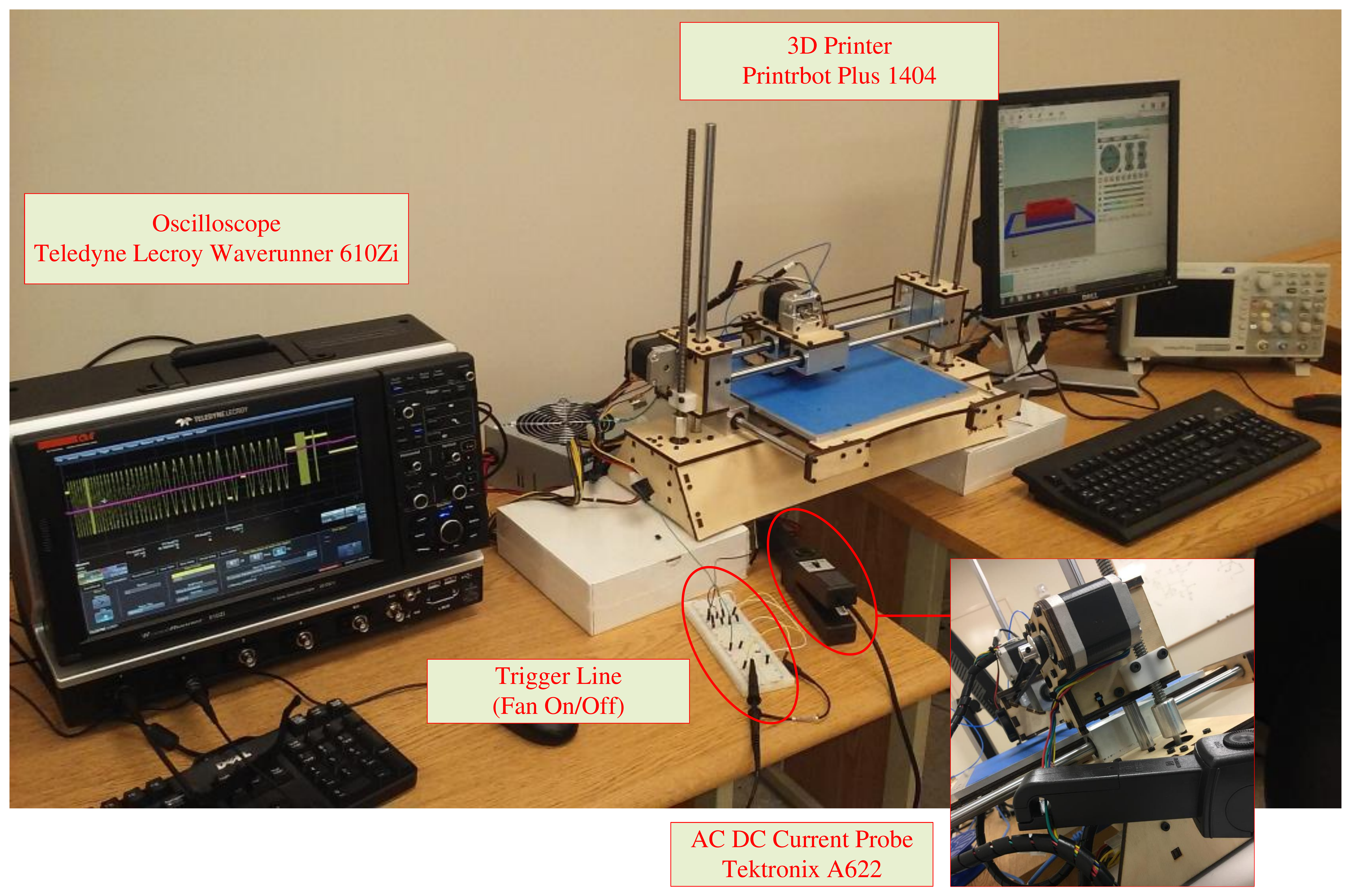}
	\caption{Experimental Environment}
	\label{fig:ExperimentalEnvironment}
\end{figure}

Figure~\ref{fig:ExperimentalEnvironment} presents our experimental setup. 
%http://archives.sensorsmag.com/articles/0799/26/
We are measuring the current delivered by a single phase of the X, Y, and Z axis motors, along with the extruder motor.
To avoid reducing the delivered current or introducing phase shifts, we use a noninvasive device: a Tektronix A622 AC/DC current probe. 
It is connected to a Teledyne LeCroy Waverunner 610 Zi oscilloscope, running at a 25 KS/s sampling rate.
According to our preliminary evaluation, this was vastly in excess of the major frequency components, which were expected in the 0-200 Hz range. We are therefore we above the Nyquist rate for accurately sampling these signals, which is double the highest frequency component. As there were no difficulties collecting at the higher sampling rate, we did not reduce it.

To better synchronize our collected traces, we implemented a trigger signal.
We selected the 3D Printer's extruder fan, as it was not one of the actuator manipulations under test. 
The fan control line was pulled to an external resistor, and set to generate a falling edge immediately before printing the first layer. This triggered the oscilloscope to begin sampling.

The experiments were run using a Printrbot Plus 1404 desktop 3D printer, employing Fused Deposition Modeling (FDM) technology.
The printer uses Nema 17 stepper motors, rated for 4.2V and 1.5 A per phase.

A desktop PC using the Repetier-Host software controlled the print jobs. Riscure Inspector performed the data preprocessing and analytics.
%--------------------------------------------------

\subsection{Experiments Performed}

The goal of the experiments is to test the proposed solution and to evaluate its sensitivity threshold to manipulations.
While maliciously inserted print defects might be small, as discussed in the literature, creating them can introduce significant changes to the G-code. This is because a slicer will have to ``work around'' gaps or other changes. 
To produce smaller and more controllable manipulations, we have altered individual G-code commands within a single print job.

We first designed a benign object, a 10 layer cube with a honeycomb fill. For each of the X, Y, Z, and extruder motors, we collected a minimum of 10 traces of the object being printed, representing 40 total prints. These measurements established a baseline or ``golden'' measurement for the motors.

We next created manipulated copies of this object, with the following modifications:
\begin{itemize}
	\item Insertion of a new G-code command, a G0 move, in layer 7.
	\item Deletion of a G-code command present in the original STL file, a G1 move in layer 7.
	\item A reordering of two G-code commands present in the original STL file, two G1 moves in layer 7 and two more in layer 8. 
	\item Replacement of a G1 command for simultaneous movement and filament extrusion with the  G0 movement only.
\end{itemize}

We collected at least 3 traces of each malicious print per motor. A comparison against a known good trace produced the deviation over time. The detection threshold for attacks is 0.1 amps above the peak amplitude of the normal traces' standard deviation.

%--------------------------------------------------

\subsection{Experimental Results}

The experimental results are summarized in Table~\ref{table:detectability} and described below in detail. All trace captures have been smoothed by a moving average filter spanning 20 samples.

\begin{table}[tbp]
	\caption{Results of the detection method on the considered attacks.}
	\begin{center}
	\begin{tabular}{r@{\quad}llll}
	\hline\noalign{\smallskip}
	
	{} & \textsc{X Motor} & \textsc{Y Motor} & \textsc{Z Motor} & \textsc{Extruder} \\
	
	\noalign{\smallskip}
	\hline 
	\noalign{\smallskip}
	
	\textsc{Normal}  & No (correct)  & No (correct) & No (correct) & No (correct) \\ 
	\textsc{Insert}  & Yes & Yes & Visible & Visible \\
	\textsc{Delete}  & Yes & Yes & Visible & Visible \\
	\textsc{Reorder} & Yes & Yes & Visible & Visible \\[2pt]
	\textsc{Void}    & Irrelevant  & Irrelevant  & Irrelevant & Visible \\
	
	\hline
	\end{tabular}
	\end{center}
	\label{table:detectability}
\end{table}

%--------------------------------------

\subsubsection{Normal Operation Traces and Standard Deviations}

\begin{figure}[h!]
	%\centerline{\includegraphics[width=1.00\textwidth, keepaspectratio]{img/X_Normal_Zoom.jpg}}
	\centerline{\includegraphics[width=1.00\textwidth, keepaspectratio]{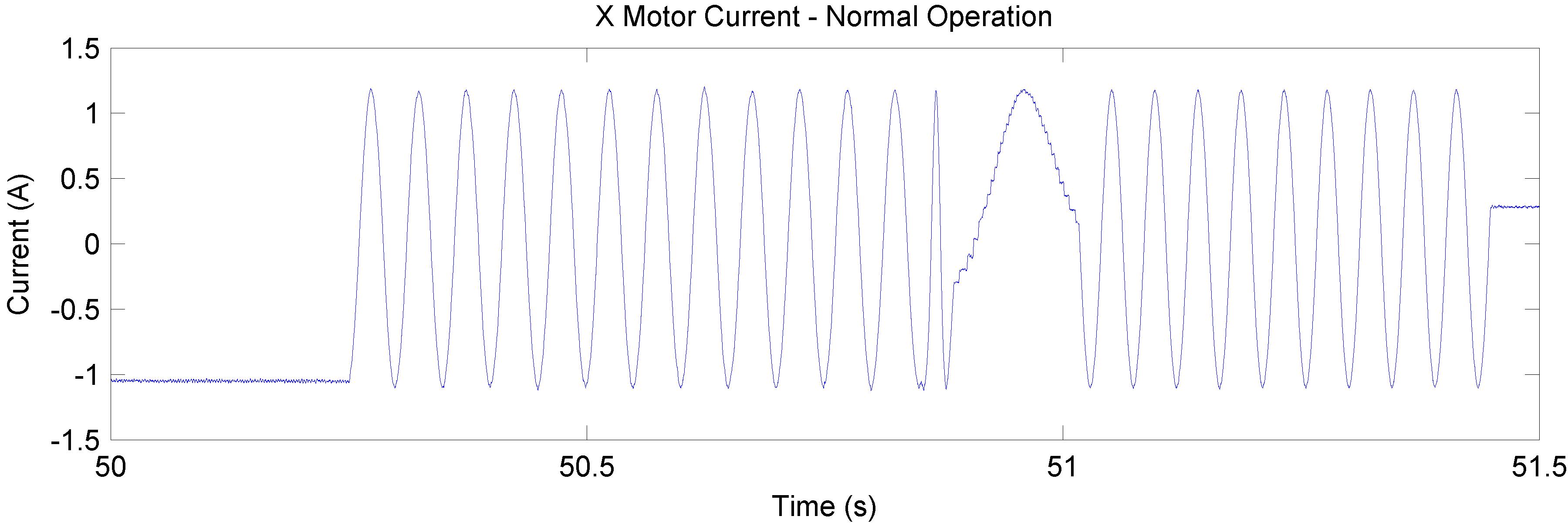}}
	\caption{Normal X motor operation.}
	\label{fig:XNormalZoom}
\end{figure}

The X and Y motors' normal operation produced a signal similar to our expectations. These motors are active throughout most of the print. In Figure \ref{fig:XNormalZoom}, a smaller timescale shows that the active sections are strongly periodic, separated by constant-level sections with high frequency noise.

\begin{figure}[h!]
	%\centerline{\includegraphics[width=1.00\textwidth, keepaspectratio]{img/X_Normal_Std_Dev.jpg}}
	\centerline{\includegraphics[width=1.00\textwidth, keepaspectratio]{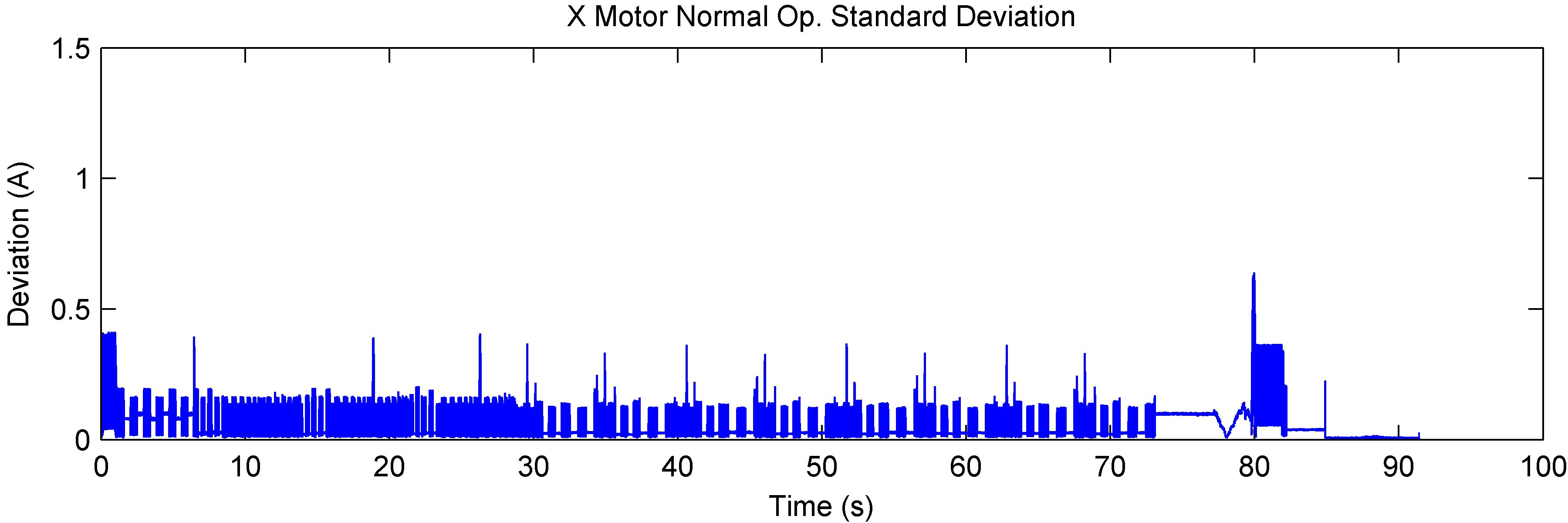}}
	\caption{Standard deviation of X traces over time.}
	\label{fig:XNormalStdDev}
\end{figure}

The standard deviation of the X and Y traces (Figure \ref{fig:XNormalStdDev}) varies over time, but remains below 0.5 A until after the print is completed at approximately 75 seconds.

\begin{figure}[h!]
	%\centerline{\includegraphics[width=1.00\textwidth, keepaspectratio]{img/Z_Normal.jpg}}
	\centerline{\includegraphics[width=1.00\textwidth, keepaspectratio]{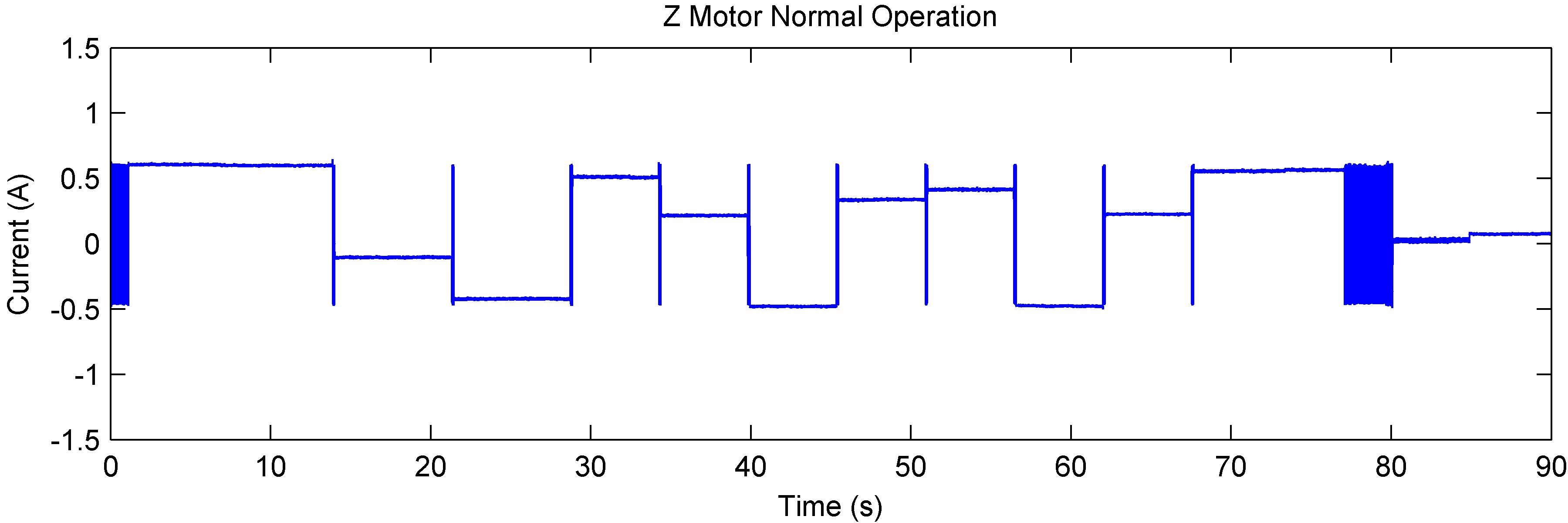}}
	\caption{Normal Z motor operation.}
	\label{fig:ZNormal}

	%\centerline{\includegraphics[width=1.00\textwidth, keepaspectratio]{img/Z_Normal_Std_Dev.jpg}}
	\centerline{\includegraphics[width=1.00\textwidth, keepaspectratio]{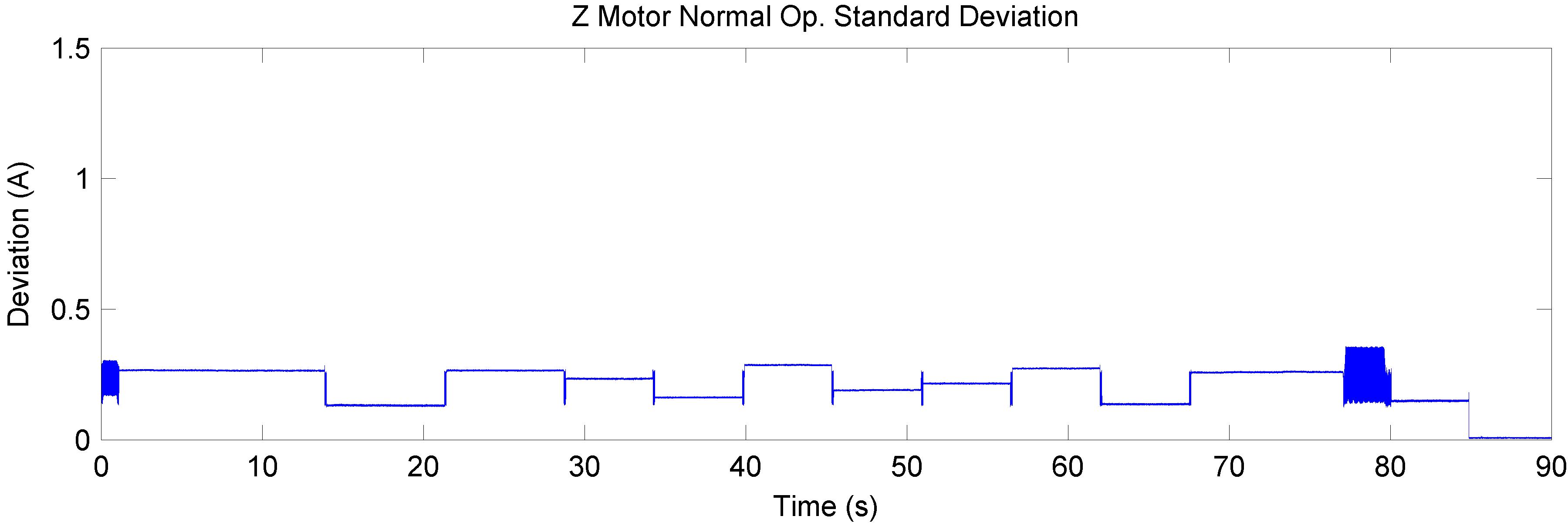}}
	\caption{Standard deviation of Z traces over time.}
	\label{fig:ZNormalStdDev}
\end{figure}

The Z motor traces show a different usage pattern. After the beginning of the print, the Z motor is active infrequently (Figure \ref{fig:ZNormal}); this matches the common-sense observation that the Z motor is used only at layer transitions. This effect also shows in the standard deviation plot (Figure \ref{fig:ZNormalStdDev}). The deviation remains constant for long periods, and is higher than the standard deviation of the X and Y motor traces while inactive. This indicates the constant current levels held between active periods are less consistent for the Z motor. As observed in the other motors (Figure \ref{fig:XNormalZoom}), the level holds where the previous periodic section ended.

\begin{figure}[h!]
	%\centerline{\includegraphics[width=1.00\textwidth, keepaspectratio]{img/E_Normal_Std_Dev.jpg}}
	\centerline{\includegraphics[width=1.00\textwidth, keepaspectratio]{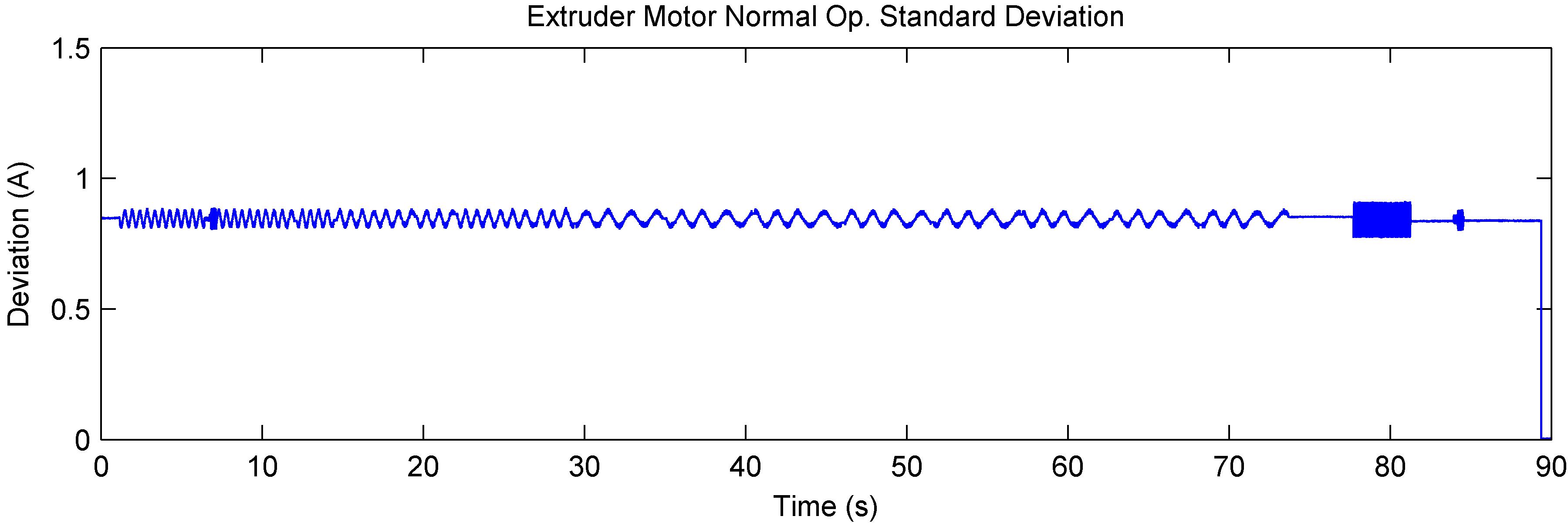}}
	\caption{Standard deviation of extruder traces over time.}
	\label{fig:ENormalStdDev}
\end{figure}

Traces from the extruder motor show lower-frequency operation with very few interruptions, but a consistently higher standard deviation (Figure \ref{fig:ENormalStdDev}).

\subsubsection{Insertion Attack}

\begin{figure}[h]
	%\centerline{\includegraphics[width=1.00\textwidth, keepaspectratio]{img/X_Insert_Full_Process.jpg}}
	\centerline{\includegraphics[width=1.00\textwidth, keepaspectratio]{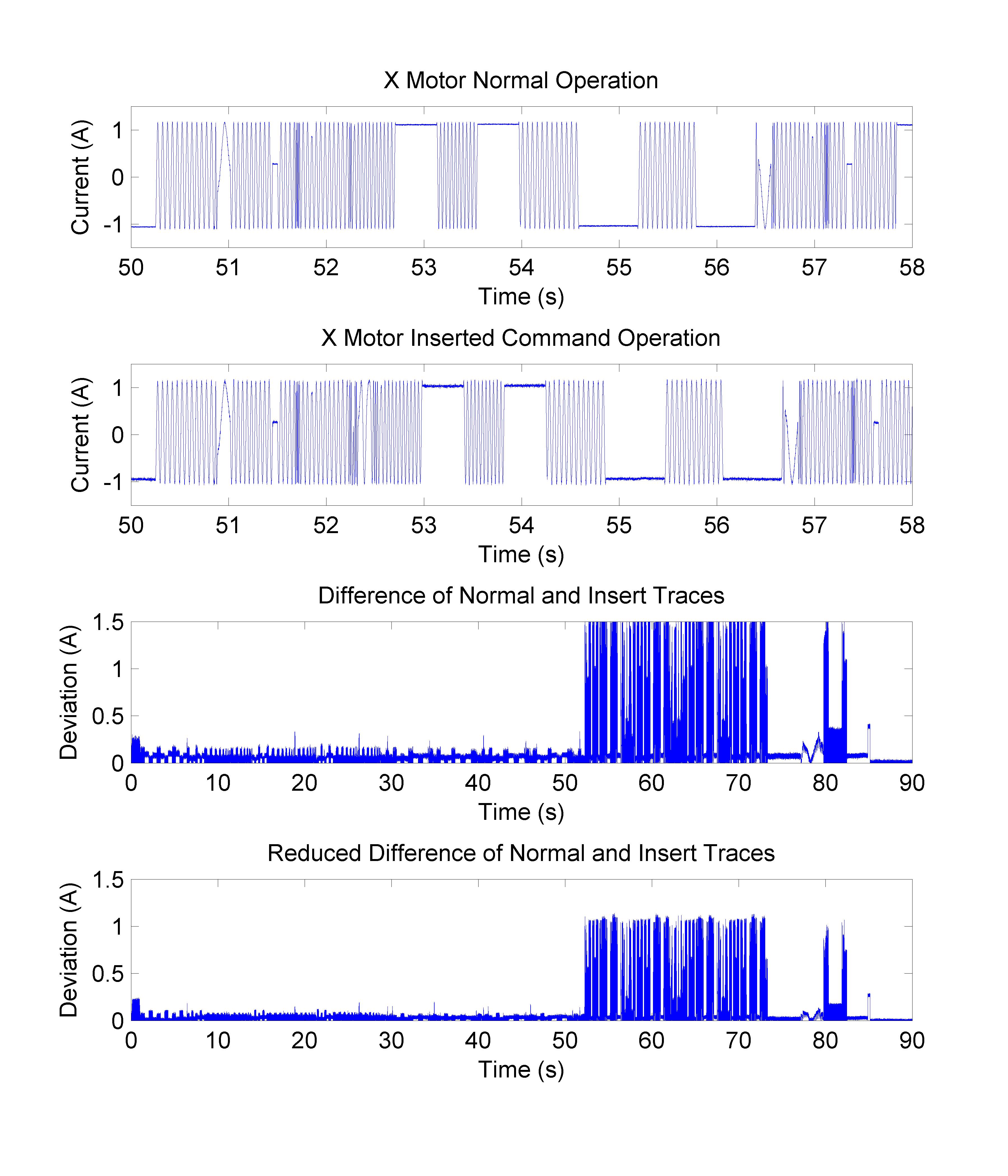}}
	\caption{The analysis process for an Insertion attack.}
	\label{fig:XFullProcess}
\end{figure}

The results of our method applied to an Insertion modified object are in Figure \ref{fig:XFullProcess}. The execution of the extra command occurs between 52 s and 53 s in the second trace; traces 1 and 2 are desynchronized by the duration of the inserted command. This appears in the difference plot as an increase in amplitude from below 0.5 A to nearly 2 A. After reducing the difference plot by subtracting the standard deviation of normal X traces, the malicious section is still clearly visible.

\begin{figure}[h]
	%\centerline{\includegraphics[width=1.00\textwidth, keepaspectratio]{img/Z_Insert_Full.jpg}}
	\centerline{\includegraphics[width=1.00\textwidth, keepaspectratio]{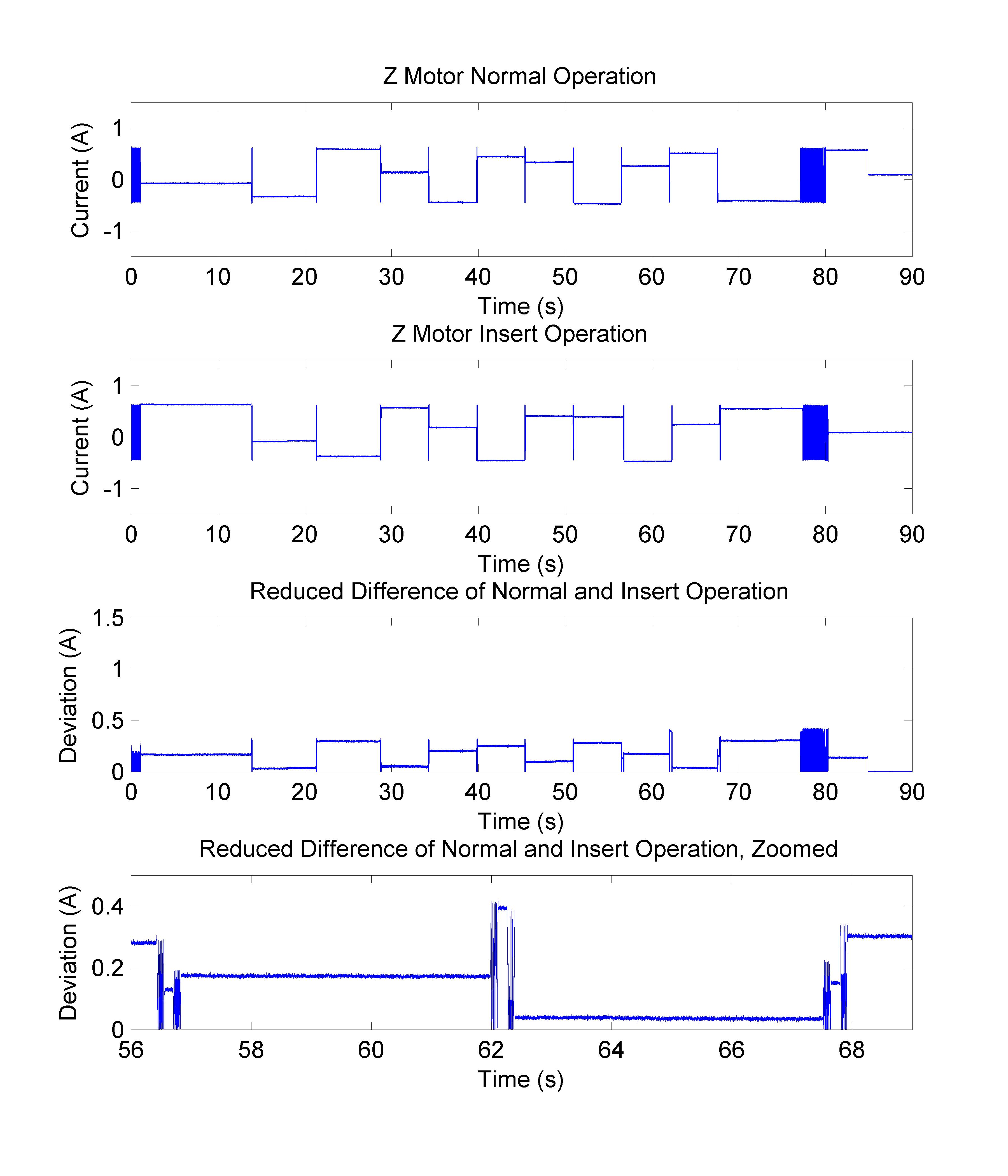}}
	\caption{Z motor behavior during an Insertion attack.}
	\label{fig:ZInsert}
\end{figure}

This attack was detected on the X and Y motors by a safe margin, as seen in Figure \ref{fig:XFullProcess}. It was not detected on the Z or Extruder motors, as seen in Figure \ref{fig:ZInsert}. The attacks were visible in the Z and Extruder traces, but did not result in larger deviations during or after the attack period.

The detection failure on the Z trace is due to the very brief active period of the Z motor. The time delay of a single command insertion is enough to misalign the motor activity, producing the duplicated spikes seen in the final plot of Figure \ref{fig:ZInsert}. Detection failure on the extruder motor is more likely due to the already high standard deviation of the normal traces (Figure \ref{fig:ENormalStdDev}). The Extruder signal shows much less synchronization than other motors. As is seen more clearly in the Void attack (Figure \ref{fig:FullVoid}), periods of inactivity or misalignment in the extruder do not show substantially higher deviation.

\subsubsection{Deletion Attack}

A Deletion attack produces similar deviations in all motors to the Insertion attack; the amount of desynchronization introduced is nearly identical. The detectability is the same in each case.

\clearpage

\subsubsection{Reorder Attack}

\begin{figure}[h!]
	%\centerline{\includegraphics[width=1.00\textwidth, keepaspectratio]{img/X_Reorder_Diff_Reduced.jpg}}
	\centerline{\includegraphics[width=1.00\textwidth, keepaspectratio]{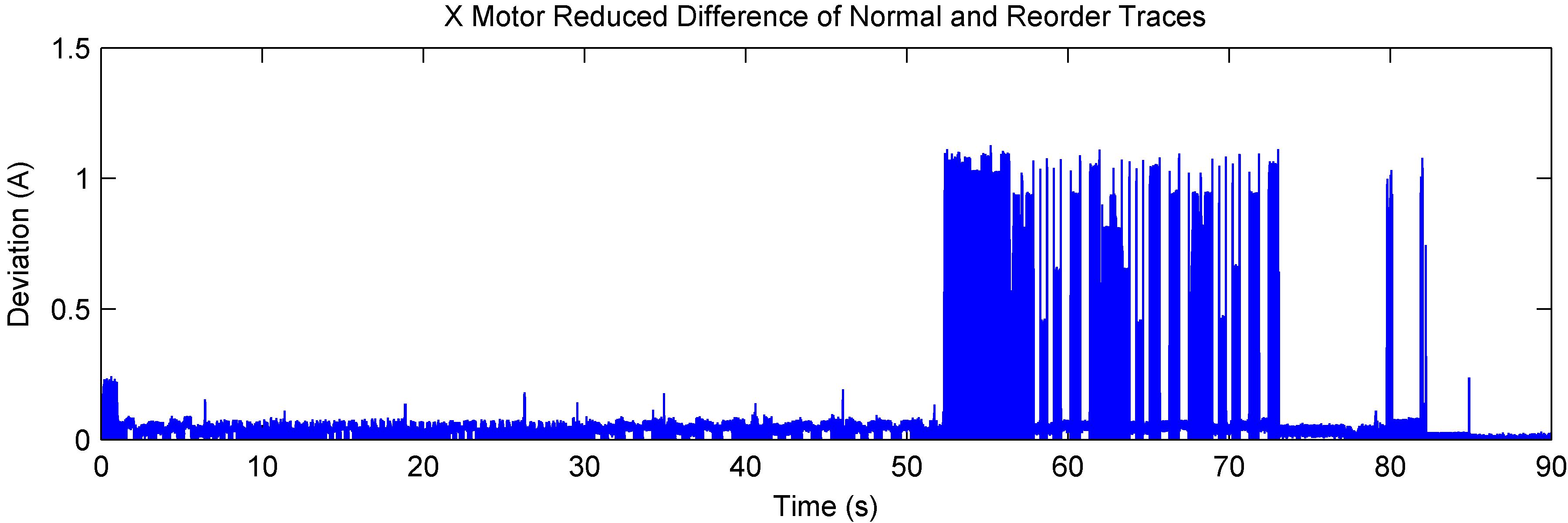}}
	\caption{X motor deviation during a Reorder Attack.}
	\label{fig:XReorderDiff}
\end{figure}

A Reorder attack produces the expected reversal of two adjacent sections of samples, but also results in a greater than average deviation after the G-code commands return to normal (Figure \ref{fig:XReorderDiff}). This may be because the reordered commands were both move commands, with a starting and ending point specified. Swapping them creates a greater distance between the endpoint of one move and the start point of the other, causing a delay.

Detectability of the attack is identical to the Insertion and Deletion attacks. The high-activity X and Y motors show significant deviations due to the misaligned active sections. The Z motor also shows misalignment, but the short activity periods result in separate, smaller deviation spikes. The extruder motor deviation is even smaller than in the Insertion or Deletion attacks, but the inactive time is still visible.

\subsubsection{Void Attack}

\begin{figure}[h]
	%\centerline{\includegraphics[width=1.00\textwidth, keepaspectratio]{img/Full_Void.jpg}}
	\centerline{\includegraphics[width=1.00\textwidth, keepaspectratio]{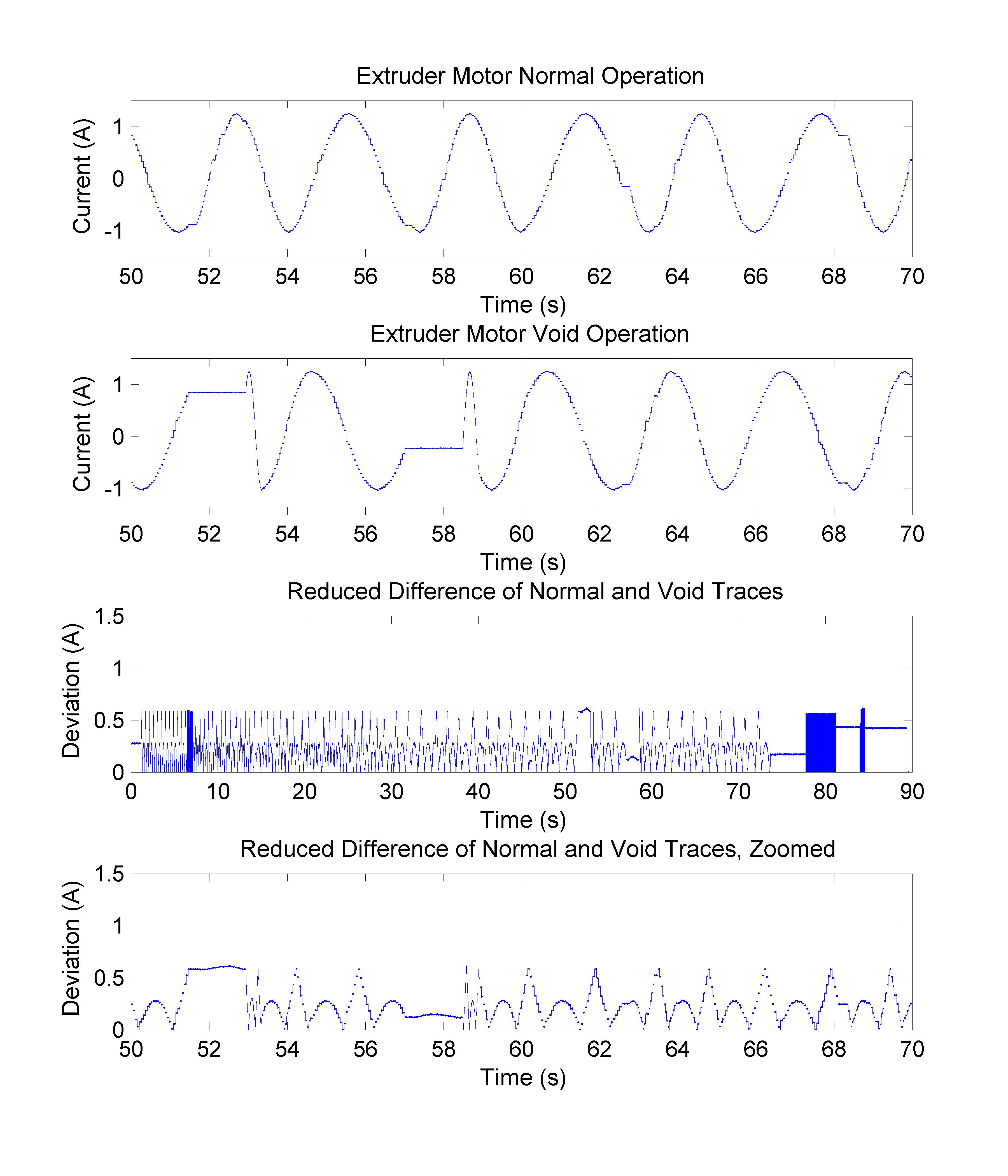}}
	\caption{Extruder motor operation during a Void print.}
	\label{fig:FullVoid}
\end{figure}

During the Void attack, the inactive time due to the modification, while similar in length to the X and Y motors in other attacks, is only a fraction of a single cycle (Figure \ref{fig:FullVoid}). The deviation across these sections is not substantially higher than the maximum deviation throughout the print.

This attack is not detected in the X, Y, and Z traces; the deviation is below the threshold set by the standard deviation for each. It was also not detected in the extruder trace, but was clearly visible.

The X, Y, and Z detectability is expected for this attack; the modified code did not alter the movement of the print head or bed z-level at all. The inability of the method to detect changes in the extruder trace is due to the already high deviation of these traces, and that the attacks do not significantly misalign the traces. In fact, the extruder activity was almost 180$^{\circ}$ out of phase before the attack section. The break in periodicity for the duration of the attack is clear, however (Figure \ref{fig:FullVoid}, final plot).

\clearpage

% ==================================================================
% === EXPERIMENTAL EVALUATION
% ==================================================================

\section {Discussion}
\label{sec:discussion}

While the proposed approach has shown impressive results in the tested experimental setup, two questions should be discussed:
(i) limitations of the approach, and
(ii) its applicability to metal AM systems.

\subsection{Approach Limitations}

Detection of effects on the Z-axis and extruder motors was not successful with an absolute-deviation-based threshold, although the effects were visible in the current traces. As the deviation is a result of comparing out of phase periodic signals, a frequency-based measure may better detect these effects while still detecting the effects on the X and Y motors.

It is possible that alterations to the G-code can produce a deviation in the trace that is not detected by our method, which still results in a malicious effect. For example, an accumulation of slight modifications to the movement speed of the X, Y, and Z motors may eventually deposit significantly more filament; the method would need to also detect small but prolonged deviations in this case. Testing a wider range of possible G-code modifications would verify this.

Our methodology assumes an unmediated relationship between the parameters of the G-code commands and the resulting actuation signal. This is not the case in closed-loop control systems, where the actuator signal is a function of both the input commands and sensor feedback. In such cases, signals in the feedback loop would also need to be captured.

% ---------------------------------------------

\subsection{Applicability to Metal AM Systems}

While the proposed approach has shown good results for FDM technology, this technology is predominantly used with plastics, a source material that has little relevance for safety-critical applications.
Two other AM processes are dominating the field in metal additive manufacturing: Powder Bed Fusion (PBF) and Directed Energy Deposition (DED).

In PBF, a thin layer of powder is deposited in a bed and the next layer's profile is melted with a laser or electron beam.
Applying the proposed approach to a PBF system will likely require additional instrumentation of heat sources, in order to detect changes in scanning strategies: Selective Laser Sintering/Melting (SLS/SLM) might require a camera and an infrared image recognition system; monitoring Electron Beam Melting (EBM) might utilize EM emanations instead.

DED systems employ a multi-axis arm, through a nozzle mounted on its end melted material is deposited onto a surface, where it solidifies. The source material is in either wire or powder form; it is melted using a laser, electron beam, or plasma arc.
The proposed approach may be directly applicable for sabotage attack detection in DED systems if the motors controlling the arm can be measured using the same techniques.

% ==================================================================
% === CONCLUSION
% ==================================================================
\section{Conclusion and Future Work}

Additive Manufacturing (AM), a.k.a. 3D Printing, is increasingly adopted around the world and used to produce functional parts of safety-critical systems.
Because of AM's dependence on computerization, there is a growing concern that the AM process can be tampered with, in order to sabotage a part's mechanical properties. 
To address this threat, we proposed a novel approach for detecting sabotage attacks in manufacturing systems. 
Our approach is based on the continuous monitoring of current supplied to all actuators during the manufacturing process and detecting anomalies compared to a provable benign process. 

The proposed approach has numerous advantages: (i) it is non-invasive in a time-critical process (ii) it can be retrofitted in legacy systems, and (iii) it is air-gapped from the computerized components involved in AM process, increasing the difficulty of a simultaneous compromise. 
We have evaluated the proposed approach on a desktop 3D Printer employing Fused Deposition Modeling (FDM) technology. 
We monitored power supply to four motors: the X/Y/Z-axis motors, and the filament extrusion motor.
Our results show that the insertion, deletion, and reordering of individual G-code movement commands can be detected with 100\% precision through the X and Y motors. Modifications to extrusion rate are visible, but not detectable with the current method.

This method can detect attacks depending on these modifications, which includes many of the void-insertion attacks discussed in the literature and any attacks relying on modifications to a 3D model file.

In our future work, we plan to overcome the identified limitations, including restriction to open-loop AM systems, switching to a frequency-based deviation measure, and accounting for the gradual accumulation of deviations. We will also test the method against other FDM printers, and adapt it for other printing technologies, such as Powder Bed Fusion and Directed Energy Deposition. 
The demonstrated anomaly detection performance and the potential applicability to metal AM systems makes the proposed approach an important milestone to ensure AM security in safety-critical systems.

\bibliographystyle{splncs03}
\bibliography{PCb-SabotageAttackDetection_Refs}

\begin{thebibliography}{10}
\providecommand{\url}[1]{\texttt{#1}}
\providecommand{\urlprefix}{URL }

\bibitem{agrawal2007trojan}
Agrawal, D., Baktir, S., Karakoyunlu, D., Rohatgi, P., Sunar, B.: Trojan
  detection using ic fingerprinting. In: Security and Privacy, 2007. SP'07.
  IEEE Symposium on. pp. 296--310. IEEE (2007)

\bibitem{belikovetsky2017dr0wned}
Belikovetsky, S., Yampolskiy, M., Toh, J., Elovici, Y.: dr0wned-cyber-physical
  attack with additive manufacturing. arXiv preprint arXiv:1609.00133  (2016)

\bibitem{chhetri2016kcad}
Chhetri, S.R., Canedo, A., Al~Faruque, M.A.: Kcad: kinetic cyber-attack
  detection method for cyber-physical additive manufacturing systems. In:
  Proceedings of the 35th International Conference on Computer-Aided Design.
  p.~74. ACM (2016)

\bibitem{chhetri2017cross}
Chhetri, S.R., Wan, J., Al~Faruque, M.A.: Cross-domain security of
  cyber-physical systems. In: Design Automation Conference (ASP-DAC), 2017 22nd
  Asia and South Pacific. pp. 200--205. IEEE (2017)

\bibitem{do2016data}
Do, Q., Martini, B., Choo, K.K.R.: A data exfiltration and remote exploitation
  attack on consumer 3d printers. IEEE Transactions on Information Forensics
  and Security  11(10),  2174--2186 (2016)

\bibitem{ey2016how}
{Ernst \& Young}: {How will 3D printing make your company the strongest link in
  the value chain?} Tech. rep. (2016),
  \url{http://www.ey.com/Publication/vwLUAssets/ey-global-3d-printing-report-2016-full-report/$FILE/ey-global-3d-printing-report-2016-full-report.pdf}

\bibitem{GE2015faa}
{GE Reports}: {The FAA Cleared The First 3D Printed Part To Fly In A Commercial
  Jet Engine From GE}. Tech. rep. (2015),
  \url{http://www.gereports.com/post/116402870270/the-faa-cleared-the-first-3d-printed-part-to-fly/}

\bibitem{khajavi2014additive}
Khajavi, S.H., Partanen, J., Holmstr{\"o}m, J.: Additive manufacturing in the
  spare parts supply chain. Computers in industry  65(1),  50--63 (2014)

\bibitem{monroy2013study}
Monroy, K., Delgado, J., Ciurana, J.: Study of the pore formation on cocrmo
  alloys by selective laser melting manufacturing process. Procedia Engineering
   63,  361--369 (2013)

\bibitem{moore2017implications}
Moore, S.B., Glisson, W.B., Yampolskiy, M.: Implications of malicious 3d
  printer firmware. In: Proceedings of the 50th Hawaii International Conference
  on System Sciences (2017)

\bibitem{pope2016hazard}
Pope, G., Yampolskiy, M.: {A Hazard Analysis Technique for Additive
  Manufacturing}. In: Better Software East Conference (2016)

\bibitem{santos2006rapid}
Santos, E.C., Shiomi, M., Osakada, K., Laoui, T.: Rapid manufacturing of metal
  components by laser forming. International Journal of Machine Tools and
  Manufacture  46(12),  1459--1468 (2006)

\bibitem{schwindling2015two}
Schwindling, F.S., Seubert, M., Rues, S., Koke, U., Schmitter, M., Stober, T.:
  Two-body wear of cocr fabricated by selective laser melting compared with
  different dental alloys. Tribology Letters  60(2), ~25 (2015)

\bibitem{slaughter2017ensure}
Slaughter, A., Yampolskiy, M., Matthews, M., King, W.E., Guss, G., Elovici, Y.:
  How to ensure bad quality in metal additive manufacturing: In-situ infrared
  thermography from the security perspective. In: Proceedings of the 12th
  International Conference on Availability, Reliability and Security. No.~78,
  ACM (2017)

\bibitem{sturm2014cyber}
Sturm, L., Williams, C., Camelio, J., White, J., Parker, R.: Cyber-physical
  vunerabilities in additive manufacturing systems. Context  7, ~8 (2014)

\bibitem{strum2017insitu}
Sturm, L., Albakri, M., Williams, C.B., Tarazaga, P.: In-situ detection of
  build defects in additive manufacturing via impedance-based monitoring

\bibitem{wohlers2016report}
Wohlers, T.: Wohlers report 2016 3d printing and additive manufacturing state
  of the industry annual worldwide progress report (2016),
  \url{www.wohlersassociates.com}

\bibitem{wu2017detecting}
Wu, M., Song, Z., Moon, Y.B.: Detecting cyber-physical attacks in
  cybermanufacturing systems with machine learning methods. Journal of
  Intelligent Manufacturing pp. 1--13 (2017)

\bibitem{yampolskiy2013taxonomy}
Yampolskiy, M., Horvath, P., Koutsoukos, X.D., Xue, Y., Sztipanovits, J.:
  Taxonomy for description of cross-domain attacks on cps. In: Proceedings of
  the 2nd ACM international conference on High confidence networked systems.
  pp. 135--142. ACM (2013)

\bibitem{yampolskiy2015security}
Yampolskiy, M., Schutzle, L., Vaidya, U., Yasinsac, A.: Security challenges of
  additive manufacturing with metals and alloys. In: Critical Infrastructure
  Protection IX, pp. 169--183. Springer (2015)

\bibitem{yampolskiy2016using}
Yampolskiy, M., Skjellum, A., Kretzschmar, M., Overfelt, R.A., Sloan, K.R.,
  Yasinsac, A.: {Using 3D Printers as Weapons}. International Journal of
  Critical Infrastructure Protection  14,  58--71 (2016)

\bibitem{zeltmann2016manufacturing}
Zeltmann, S.E., Gupta, N., Tsoutsos, N.G., Maniatakos, M., Rajendran, J.,
  Karri, R.: Manufacturing and security challenges in 3d printing. JOM pp.
  1--10 (2016)

\end{thebibliography}

\end{document}